\documentclass[aps,pra,twocolumn,showpacs,amsmath,amssymb,preprintnumbers,superscriptaddress,10pt]{revtex4-1}

\usepackage{bm}
\usepackage{graphicx}
\usepackage{SIunits}
\usepackage{cleveref}
\crefname{section}{Sec.}{Secs.}
\Crefname{section}{Section}{Sections}
\usepackage{color,colortbl}
\definecolor{lightgray}{rgb}{0.85,0.85,0.85}
\usepackage{multirow}
\usepackage{array}
\usepackage[normalem]{ulem}
\newcolumntype{L}[1]{>{\raggedright\let\newline\\\arraybackslash\hspace{0pt}}m{#1}}
\newcolumntype{C}[1]{>{\centering\let\newline\\\arraybackslash\hspace{0pt}}m{#1}}
\newcolumntype{R}[1]{>{\raggedleft\let\newline\\\arraybackslash\hspace{0pt}}m{#1}}

\definecolor{pink}{RGB}{255,0,255}

\definecolor{purple}{rgb}{0.6,0,1}

\definecolor{ss_color}{rgb}{1,0,0}

\definecolor{ngreen}{rgb}{0.2,0.6,0.2}

\definecolor{ngreen2}{rgb}{0.2,0.6,0.6}

\begin{document}

\title{Eavesdropper's ability to attack a free-space quantum-key-distribution receiver in atmospheric turbulence}

\author{Poompong~Chaiwongkhot}
\email{poompong.ch@gmail.com}
\affiliation{Institute for Quantum Computing, University of Waterloo, Waterloo, ON, N2L~3G1 Canada}
\affiliation{Department of Physics and Astronomy, University of Waterloo, Waterloo, ON, N2L~3G1 Canada}

\author{Katanya B. Kuntz}
\email{katanyab@gmail.com}
\affiliation{Institute for Quantum Computing, University of Waterloo, Waterloo, ON, N2L~3G1 Canada}
\affiliation{Department of Physics and Astronomy, University of Waterloo, Waterloo, ON, N2L~3G1 Canada}

\author{Yanbao~Zhang}
\affiliation{Institute for Quantum Computing, University of Waterloo, Waterloo, ON, N2L~3G1 Canada}
\affiliation{Department of Physics and Astronomy, University of Waterloo, Waterloo, ON, N2L~3G1 Canada}
\affiliation{NTT Basic Research Laboratories, NTT Corporation, Atsugi, Kanagawa, Japan}
\affiliation{NTT Research Center for Theoretical Quantum Physics, NTT Corporation, Atsugi, Kanagawa, Japan}

\author{Anqi~Huang}
\affiliation{Institute for Quantum Information \& State Key Laboratory of High Performance Computing, College of Computer, National University of Defense Technology, Changsha 410073, People's Republic of China}
\affiliation{Institute for Quantum Computing, University of Waterloo, Waterloo, ON, N2L~3G1 Canada}
\affiliation{\mbox{Department of Electrical and Computer Engineering, University of Waterloo, Waterloo, ON, N2L~3G1 Canada}}

\author{Jean-Philippe~Bourgoin}
\affiliation{Aegis Quantum, Waterloo, ON, Canada}
\affiliation{Institute for Quantum Computing, University of Waterloo, Waterloo, ON, N2L~3G1 Canada}
\affiliation{Department of Physics and Astronomy, University of Waterloo, Waterloo, ON, N2L~3G1 Canada}

\author{Shihan~Sajeed}
\affiliation{Department of Electrical and Computer Engineering, University of Toronto, M5S~3G4, Canada}
\affiliation{Institute for Quantum Computing, University of Waterloo, Waterloo, ON, N2L~3G1 Canada}
\affiliation{\mbox{Department of Electrical and Computer Engineering, University of Waterloo, Waterloo, ON, N2L~3G1 Canada}}

\author{Norbert~L\"{u}tkenhaus}
\affiliation{Institute for Quantum Computing, University of Waterloo, Waterloo, ON, N2L~3G1 Canada}
\affiliation{Department of Physics and Astronomy, University of Waterloo, Waterloo, ON, N2L~3G1 Canada}

\author{Thomas~Jennewein}
\affiliation{Institute for Quantum Computing, University of Waterloo, Waterloo, ON, N2L~3G1 Canada}
\affiliation{Department of Physics and Astronomy, University of Waterloo, Waterloo, ON, N2L~3G1 Canada}
\affiliation{Quantum Information Science Program, Canadian Institute for Advanced Research, Toronto, ON, M5G~1Z8 Canada}

\author{Vadim~Makarov}
\affiliation{Russian Quantum Center, Skolkovo, Moscow 143025, Russia}
\affiliation{\mbox{Shanghai Branch, National Laboratory for Physical Sciences at Microscale and CAS Center for Excellence in} \mbox{Quantum Information, University of Science and Technology of China, Shanghai 201315, People's Republic of China}}
\affiliation{NTI Center for Quantum Communications, National University of Science and Technology MISiS, Moscow 119049, Russia}
\affiliation{Department of Physics and Astronomy, University of Waterloo, Waterloo, ON, N2L~3G1 Canada}

\date{\today}

\begin{abstract}
The ability of an eavesdropper (Eve) to perform an intercept-resend attack on a free-space quantum key distribution (QKD) receiver by precisely controlling the incidence angle of an attack laser has been previously demonstrated. However, such an attack could be ineffective in the presence of atmospheric turbulence due to beam wander and spatial mode aberrations induced by the air's varying index of refraction. We experimentally investigate the impact turbulence has on Eve's attack on a free-space polarization-encoding QKD receiver by emulating atmospheric turbulence with a spatial light modulator. Our results identify how well Eve would need to compensate for turbulence to perform a successful attack by either reducing her distance to the receiver, or using beam wavefront correction via adaptive optics. Furthermore, we use an entanglement-breaking scheme to find a theoretical limit on the turbulence strength that hinders Eve's attack.
\end{abstract}

\maketitle

\section{Introduction}

Quantum key distribution (QKD) allows two distant parties to exchange secret keys with -- in theory -- unconditional security \cite{bennett1984,ekert1991}. However in practice, a QKD system is often not perfect, and unconditional security cannot be guaranteed. Any imperfections in the physical implementation of a QKD scheme can lead to side-channels that could be exploited by an eavesdropper (Eve) and compromise security \cite{vakhitov2001,makarov2006,qi2007,lamas-linares2007,lydersen2010a,xu2010,gerhardt2011,weier2011,jouguet2013,sajeed2015,sajeed2015a, makarov2016,sajeed2016,meda2017,sajeed2017}. Therefore, it is of utmost importance to perform security evaluations of practical systems, i.e.,\ scrutinize vulnerabilities, determine useful testing methodologies and assess the risk to formulate countermeasures for preventing successful attacks. 

A widely studied implementation of QKD utilizes free-space communication between two parties (Alice and Bob) through the atmosphere  \cite{ursin2007,erven2008,peloso2009, nauerth2013, liao2017,pugh2017, yin2017}, which allows for long distance point-to-point links on the order of 100km. This communication distance can be extended even further to the global scale by introducing satellite-based QKD systems \cite{ursin2009,bourgoin2013,yin2013,pugh2017,liao2017,yin2017,vallone2015,casado2018}. However, free-space communication can be vulnerable to an eavesdropper attack, such as when Eve precisely controls the incidence angle of an attack laser directed at Bob's QKD receiver. Directing a laser in this way can induce a change in the measurement efficiencies of one (or more) detection channels, which enables Eve to do an intercept-resend (IR) attack that may compromise the system's security \cite{sajeed2015a,rau2015}.

The success of this spatial mode attack depends on the eavesdropper's ability to precisely maintain specific beam angles to a free-space QKD receiver, which attacks different detection channels. Atmospheric turbulence could compromise or even prevent such an attack as turbulence causes a beam to randomly wander along its trajectory, as well as inducing various optical aberrations such as astigmatism, defocus, coma, etc. Stronger turbulence conditions result in a larger variance in the amount of beam wander \cite{Tyson2010}. Consideration of these physical limitations on Eve is not usually included in the theoretical security analysis of a system, but can be useful to verify whether an attack is feasible under more realistic conditions.

In this paper, we experimentally determine the minimum strength of atmospheric turbulence that could prevent a successful attack on our free-space polarization-based QKD receiver by emulating atmospheric turbulence using a phase-only spatial light modulator (SLM). Since there are limitations on how well adaptive optics can correct for turbulence, our paper explores to what level Eve must correct her attack beam to still be successful \cite{li2004,lee2006}. We assume that the sender (Alice) and the receiver (Bob) only monitor the total count rates (as opposed to the rates of individual channels), and that they use a non-decoy state BB84Bennett-Brassard 1984 (BB84) protocol \cite{bennett1984}. We also assume that Eve has access to a weak coherent pulse source and state of the art photo-detectors, and does not have a quantum repeater. Furthermore, we assume that Eve cannot replace the quantum channel with a lossless channel. We find that an attack on our free-space receiver could still succeed if Eve can correct the tip-tilt mode for turbulence as strong as $r_0 = 1.53~\centi\meter$ (assuming an initial beam diameter of $20~\centi\meter$), where $r_0$ is the atmospheric coherence length. This result defines an ``unsafe radius" of $543~\meter$ around Bob's receiver in typical sea level turbulence conditions where Eve's attack could be successful if done within this radius. 

First we discuss our SLM setup used to emulate atmospheric turbulence, and how we verified its accuracy and reproducibility in \cref{sec:TurbEmu}. Then we describe the components and operation of our free-space polarization-based QKD receiver under test in \cref{sec:qkd_system}. In \cref{sec:mismatch,sec:attack}, we discuss the results from spatial mode attacks performed in various turbulence strengths, following a similar procedure to Sajeed \textit{et al.} in Ref.~\onlinecite{sajeed2015a}. Finally, in \cref{sec:theory} we discuss an entanglement breaking scheme proposed by Zhang \textit{et al.} in Ref.~\onlinecite{zhang2017}, to theoretically verify if there exists an attack strategy for Eve, even if Alice and Bob know about their detection efficiency mismatch, and monitor the statistics of all possible detection outcomes. We conclude in \cref{sec:conclusion}.

\section{Turbulence emulator} 
\label{sec:TurbEmu}

\begin{figure*}
\includegraphics[width=130mm]{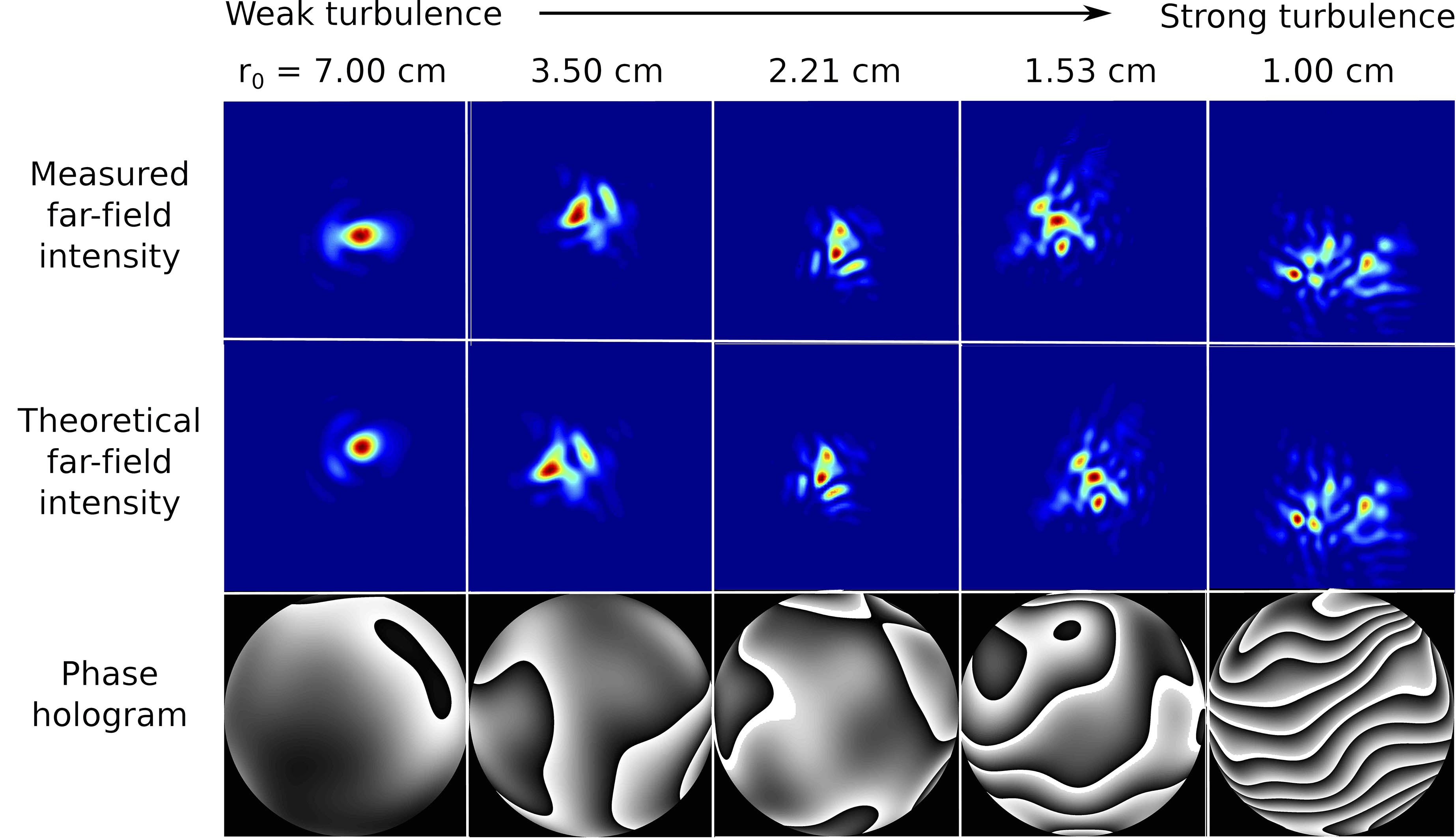}
\caption{Comparison between measured and theoretical far-field intensity distributions of a laser beam corresponding to one of 29 SLM phase holograms per turbulence strength ($r_0$) for a beam with $D = 20~\centi\meter$ and $\lambda = 532~\nano\meter$. The greyscale in the holograms represents a $0$ to $2\pi$ phase range. The results show our SLM setup accurately emulates a range of turbulence strengths.}
\label{fig:TurbDataFF}
\end{figure*}

\begin{figure}
\includegraphics{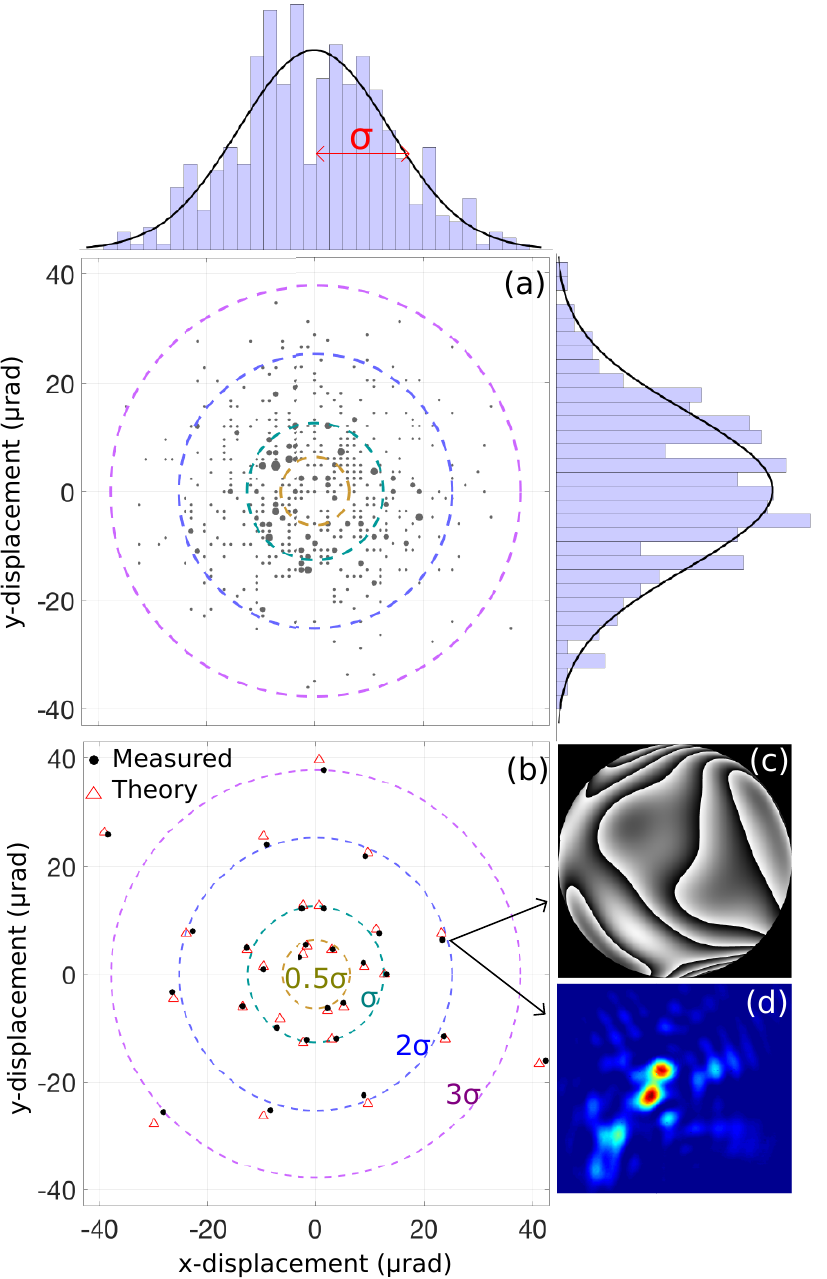}
\caption{Turbulence emulator characterization for $r_0 = 1.00~\centi\meter$, $D = 20~\centi\meter$, and $\lambda = 532~\nano\meter$. (a)~Simulated centroid displacements corresponding to 500 phase holograms ($\sigma$ is the two-axis standard deviation). The diameter of each data point is proportional to the count frequency. The centroid displacement distribution is normally distributed along both axes in agreement with \cref{eq:alpha}. (b)~Comparison between measured and simulated centroid displacements for a subset of 29 holograms. This subset was chosen to represent the normal statistical distribution of the 500-hologram set. The measured values are within error of most theoretical predictions (error bars for measured data are represented by diameter of data points). (c)~Phase hologram and (d)~far-field intensity distribution corresponding to one centroid data point.}
\label{fig:TurbDataCOM}
\end{figure}

We use a phase-only SLM to emulate a turbulent QKD channel in the laboratory. One advantage of using a SLM as opposed to performing the experiment outdoors is the ability to generate a range of turbulence strengths, from weak upper atmosphere to stronger sea level conditions. In addition, by performing our experiment in a laboratory, we are immune to the unpredictability of an outdoor environment, allowing us to repeat the same attack angles on our free-space QKD receiver under reproducible turbulence conditions.

Our model uses the `thin phase screen approximation' which emulates turbulence using a single random phase screen in the aperture of the receiver, as opposed to requiring two holograms to model multiple parameters that incorporate both phase and amplitude variations \cite{rodenburg2014}. We assume that Eve's laser can mimic the intensity variations caused by turbulence (scintillation) \cite{erven2012}. Note that the absence of these fluctuations could arouse Alice and Bob's suspicion of an eavesdropper in the channel, although fluctuations on the time scale of scintillation at 1s or less are rarely monitored in practice.

In order to reproduce the random statistics of turbulence, we load a series of $29$ phase maps per turbulence strength on the SLM to distort the optical wavefront. The strength of the turbulence is completely characterized by the ratio of the initial beam diameter, $D$, to the atmospheric coherence length, $r_0$; turbulence dominates over diffractive effects when $D/r_0 \gg1$. 

We generate our phase holograms based on the well-known Kolmogorov model \cite{andrews1998} that uses a weighted superposition of Zernike polynomials for the basis-set \cite{noll1976}. There are several advantages to using Zernike polynomials to generate the holograms as their weights can be analytically calculated based on the turbulence strength \cite{burger2008}. Furthermore, Zernike polynomials directly relate to known optical aberrations, such as tip-tilt, defocus, astigmatism, coma, etc. Therefore, it is straightforward to characterize the SLM's ability to reliably and precisely emulate atmospheric turbulence by comparing calculated Zernike polynomial coefficients to those reconstructed by a measurement device, such as a wavefront sensor.

The radial phase function $\phi(\rho,\theta)$ that describes each hologram is given by a weighted sum of several Zernike polynomials as $\phi(\rho,\theta) = \sum_i c_i Z_i$, where $Z_i$ and $c_i$ are the Zernike polynomial and corresponding coefficient for the $i$th polynomial, respectively, following the Noll labeling convention and normalization constants \cite{noll1976}. We use 44 Zernike polynomials to ensure a complex spatial structure that can accurately emulate a range of atmospheric turbulence strengths. 

Based on the Kolmogorov model \cite{andrews1998,burger2008}, if we assume that the Zernike coefficients are normally distributed with mean zero, then $c_i$ are random drawings from distributions with variance $\sigma^2_{nm}$ defined as 
\begin{align}
\sigma^2_{nm} &= I_{nm}(D/r_0)^{5/3}, \label{eq:sigma} \\ 
r_0 &= 1.68(C_n^2Lk^2)^{-3/5}, \nonumber \\
I_{nm} &= \frac{0.15337(-1)^{n-m}(n+1)\Gamma(14/3)\Gamma(n-5/6)}{\Gamma(17/6)^2\Gamma(n+23/6)}, \nonumber
\end{align} 
where $C_n^2$ is the refractive-index structure constant of the atmosphere, $L$ is the path length through the turbulent atmosphere that has a constant $C_n^2$, $k = 2\pi/\lambda$, $\lambda$ is the laser wavelength, and $\Gamma$ is the Gamma function. The indices $n$ and $m$ are related to the Zernike polynomial order following the Noll labeling convention, where $n\geq |m|$ and $n-m$ is even \cite{noll1976}. We note that the subscript ``n'' of $C_n^2$ is not related to the index ``n'' used in the Zernike polynomials, but instead to the refractive index of the atmosphere. A single value of $C_n^2$ is used when calculating $\sigma^2_{nm}$ over each of the $n$ and $m$ indices for each atmospheric strength modelled. A large $C_n^2$ (small $r_0$) value corresponds to stronger atmospheric turbulence. An example of stronger turbulent conditions that could be found at sea level corresponds to $r_0 = 1.00~\centi\meter$ over $L = 1~\kilo\meter$ for $D = 20~\centi\meter$ at $\lambda = 532~\nano\meter$, whereas weaker conditions at high altitude corresponds to $r_0 = 7.00~\centi\meter$ \cite{andrews1998}.

Since Zernike polynomials directly relate to known optical aberrations, we can use simple equations and measurement devices (CCD camera and wavefront sensor), to independently verify and characterize our turbulence emulator. \Cref{fig:TurbDataFF} shows both the simulated and measured far-field intensity distributions of a beam after its wavefront has been distorted by the SLM hologram. Each hologram shown is one example from a set of 29 holograms per $r_0$ value used to emulate how different strengths of turbulence would affect a $20~\centi\meter$ beam at $532~\nano\meter$. We experimentally image the far-field by placing a camera in the focal plane of a lens that is located one focal length from the SLM. This arrangement maps the phase wavefront imprinted on the beam by the hologram into an intensity distribution at the camera plane. Note that we include an additional x-grating in the hologram (not shown for clarity) to spatially separate the first-order diffracted beam from the zeroth-order one, as only the first-order beam contains the pure phase wavefront. The zeroth-order (and higher-order) diffracted beams were carefully blocked shortly after the SLM.

We also verify our turbulence emulator by examining the centroid deviations caused by each hologram. This is an important characterization as beam displacements due to turbulence could dominate Eve's ability to repeatedly send a beam at precise angles to the receiver. Beam wander is the strongest effect on average as the tip-tilt coefficients ($n = 1$, $m = \pm 1$) have the largest weights overall [$I_{11} = 0.45$ from \cref{eq:sigma}], whereas defocus ($I_{20} = 0.02$) and astigmatism ($I_{22} = 0.02$) have a smaller contribution on average. Higher order aberrations can also cause centroid displacement, especially in the case of stronger turbulence. 

There is a direct relationship between the tilt angle variance of centroid displacement for two uncorrelated axes $\sigma^2$ and the turbulence strength $r_0$, which is given by \cite{Tyson2010}
\begin{equation} \label{eq:alpha}
\sigma^2 = 0.364\bigg(\frac{D}{r_0}\bigg)^{5/3}\bigg(\frac{\lambda}{r_0}\bigg)^{5/3}.
\end{equation}
Since this equation is independent of the method used to emulate turbulence, we can verify whether the 29 chosen phase holograms accurately portray the statistics of atmospheric turbulence both theoretically via computer simulations of far-field intensity distributions, and experimentally through our SLM setup. This independent verification ensures that the holograms are accurate, as well as that the SLM is correctly imprinting the phase mask onto the beam.

The centroid displacement data presented in \cref{fig:TurbDataCOM} corresponds to low-altitude sea level turbulence ($r_0 = 1.00~\centi\meter$ for a $20~\centi\meter$ beam). The simulated centroid displacements from 500 holograms are shown in \cref{fig:TurbDataCOM}(a). Each data point corresponds to a unique hologram [\cref{fig:TurbDataCOM}(c)] and far-field intensity distribution [\cref{fig:TurbDataCOM}(d)]. The simulated centroids follow a Gaussian distribution with a standard deviation $\sigma$ that is in agreement with \cref{eq:alpha}. These results confirm that the phase holograms we calculated properly emulate the statistics of low-altitude sea level turbulence, irrespective of the SLM setup. Similar tests were performed to verify the sets of holograms for each $r_0$ value tested in this experiment. 

We compare simulated and measured centroid displacements of 29 holograms per $r_0$ strength in \cref{fig:TurbDataCOM}(b). The number of holograms used in the hacking experiment was limited to reduce data acquisition time and stability issues while scanning. Therefore, we chose 29 holograms from a larger distribution of 500 to emulate each $r_0$ strength. The holograms were chosen based on their centroid displacements being approximately $0.5\sigma$, $\sigma$, $2\sigma$ and $3\sigma$ from the origin [along the dashed circles outlined in Figs.~\ref{fig:TurbDataCOM}(a) and \ref{fig:TurbDataCOM}(b)], along with one histogram with no turbulence representing $0\sigma$. The centroid results, along with the qualitative comparison between theoretical and measured far-field intensity distributions (\cref{fig:TurbDataFF}), confirmed we had excellent agreement between theory and experiment for turbulence emulated by our SLM setup. The 29th hologram always emulates $0\sigma$ displacement with no turbulence. The contribution of each of the 29 holograms to the emulated turbulence in subsequent experiments is weighted by its probability of occurrence, which follows a Gaussian distribution. This probability of occurrence is a definite integral of normalized Gaussian distribution over the annulus formed by the adjacent radii shown in \cref{fig:TurbDataCOM}(b). We refer to each annulus by the name of its inner radius, near which its holograms are located. The $0\sigma$ annulus, extending from zero (where its hologram is located) to $0.5\sigma$ radius, has the weight of 0.1175. The $0.5\sigma$ annulus has the weight of 0.2760, $1\sigma$ has the weight of 0.4712, $2\sigma$ has the weight of 0.1242, and $3\sigma$ (extending to infinity) has the weight of 0.0111. 

\begin{figure*}
\includegraphics{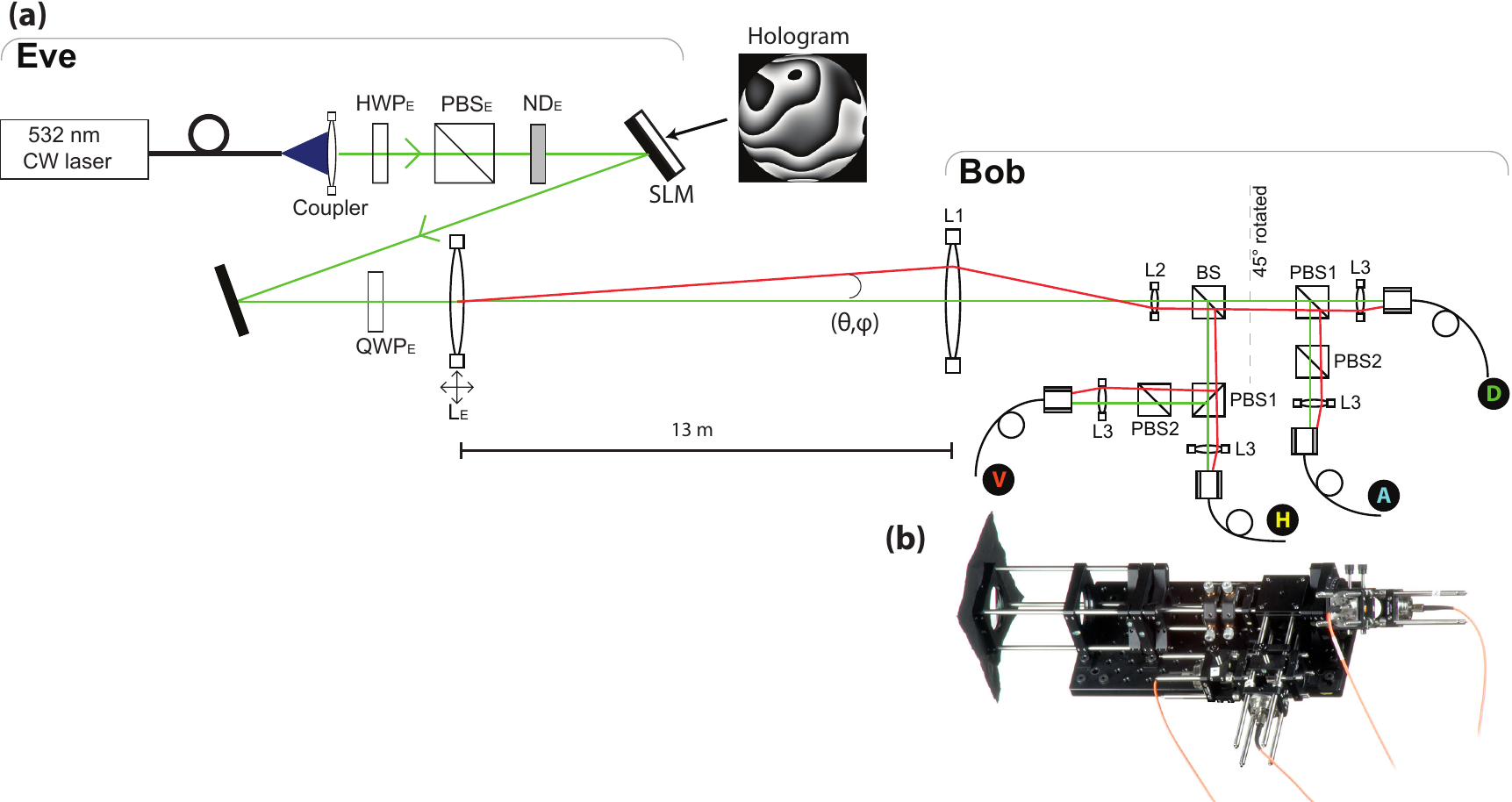}
\caption{Scanning setup. (a)~Experimental setup of our spatial mode attack in a turbulent channel, top view (drawing not to scale). The green central ray that is parallel to the optical axis denotes normal alignment of Alice's beam into Bob's receiver. The red rays show the optical path of Eve's scanning beam when tilted at an angle ($\theta,\phi$) via lens $L_E$. CW: continuous-wave; HWP: half-wave plate; QWP: quarter-wave plate; BS: beam splitter; PBS: polarization beam splitter; ND: neutral density filter; SLM: spatial light modulator; L: lens. (b)~Photograph of the actual free-space QKD receiver for detecting polarization-encoded light.}
\label{fig:setup}
\end{figure*} 

\begin{figure*}
\includegraphics{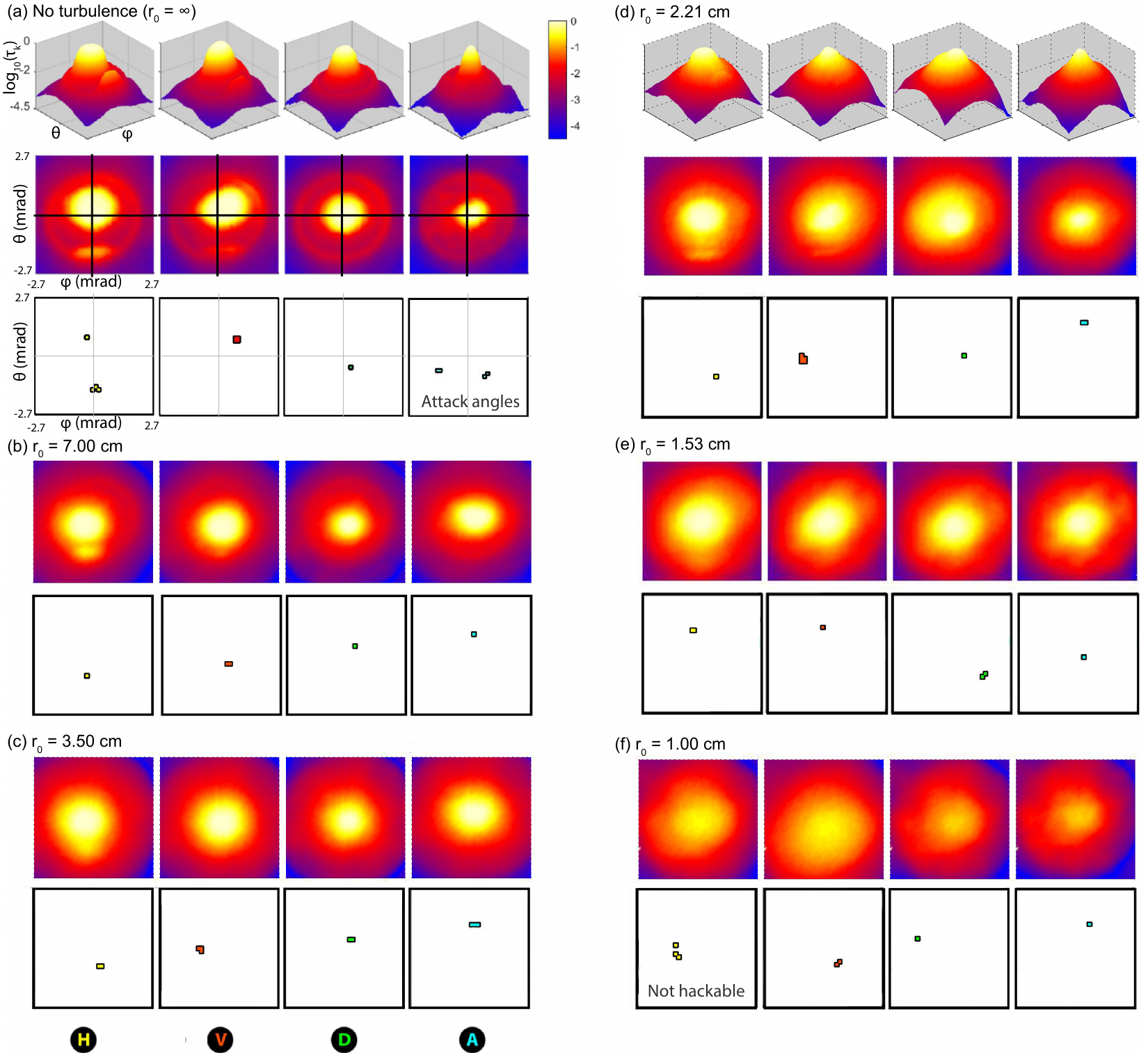}
\caption{Normalized count rates $\tau_k$ for each detector $k = \bf{H}, \bf{V}, \bf{D},\:\mathrm{or}\:\bf{A}$ at different incoming beam angles ($\theta,\phi$), and the corresponding attack angles for different turbulence strengths $r_0$. The attack angles for the four polarization detectors are shown left to right as horizontal \textbf{H} (yellow), vertical \textbf{V} (red), diagonal \textbf{D} (green), and anti-diagonal \textbf{A} (light blue). The emulated turbulence corresponds to different $r_0$ values for an initial beam diameter $D = 20~\centi\meter$ and $\lambda = 532~\nano\meter$. A smaller $r_0$ value corresponds to stronger atmospheric turbulence.}
\label{fig:scan}
\end{figure*} 

\section{Test setup for the QKD system}
\label{sec:qkd_system}

We use our turbulence emulator to study the effect of turbulence on free-space detection efficiency mismatch. Eve's experimental setup consists of two parts: the turbulence emulator (SLM) and the beam scanning unit, as shown in \cref{fig:setup}. Our source is a $532~\nano\meter$ continuous-wave laser that is first sent through a polarization beam splitter $\text{PBS}_\text{E}$ (Thorlabs CCM1-PBS251) to transmit only horizontally-polarized light to the SLM, which ensures phase-only modulation. The beam's wavefront after the SLM represents propagation through atmospheric turbulence of a particular strength. We use
a quarter-wave plate $text{QWP}_text{E}$ (Thorlabs AQWP10M-600) to rotate horizontal light to circularly polarized light to equalize the QKD receiver detector signals on the four
polarization channels. Eve's scanning lens $\text{L}_\text{E}$ is mounted on a two-axis motorized translation stage (Thorlabs MAX343/M), which scans the attack beam's angle. A half-wave plate $\text{HWP}_\text{E}$ (Thorlabs AHWP10M-600) and neutral density filter $\text{ND}_\text{E}$ (Thorlabs ND30A) are used to control Eve's intensity. Finally, the receiver is placed $13~\meter$ away from $\text{L}_\text{E}$. 

The QKD receiver under test is a prototype for a quantum communication satellite \cite{bourgoin2013}, which uses a passive basis choice to detect polarization-encoded light. Its telescope consists of a focusing lens $\text{L1}$ (diameter of $50~\milli\meter$ with a focal length $f = 250~\milli\meter$; Thorlabs AC508-250-A), and a collimating lens $\text{L2}$ (diameter of $5~\milli\meter$ with $f = 11~\milli\meter$; Thorlabs A397TM-A). The collimated beam of $\lesssim 2~\milli\meter$ diameter then passes through a 50:50 beam splitter $\text{BS}$ (custom pentaprism \cite{bourgoin2013}), and a pair of polarization beam splitters $\text{PBS1}$ and $\text{PBS2}$ (Thorlabs PBS121). The purpose of $\text{PBS2}$ is to increase the polarization extinction ratio in the reflected path from $\text{PBS1}$. The four lenses $\text{L3}$ (Thorlabs PAF-X-18-PC-A) focus the beams into four multi-mode fibers, each with a core diameter of $105~\micro\meter$ (Thorlabs M43L01), which are connected to single-photon detectors (Excelitas SPCM-AQRH-12-FC). We use one set of polarization optics and detectors to measure diagonal \textbf{D} and anti-diagonal \textbf{A} polarizations by rotating them $45\degree$ relative to the horizontal \textbf{H} and vertical \textbf{V} polarization detectors. We note that this receiver under test does not contain any active pointing system or adaptive optics.

\section{Attack using Spatial mode detection efficiency mismatch} 
\label{sec:mismatch}

This paper assumes that Alice and Bob generate a secret key using a non-decoy state BB84 protocol \cite{bennett1984}. We also make the weaker assumption presented in Ref.~\onlinecite{sajeed2015a} that they only monitor the total detection rate for evidence of Eve's attack rather than the counts of each channel. Additionally, we assume Alice and Bob also monitor only the average error rate over the four channels, and terminate the protocol if the average quantum bit error rate (QBER) over the four channels is higher than a $8\%$ threshold \cite{sajeed2015}.

The attack model we consider is an intercept-resend attack called the faked-state attack \cite{makarov2006, makarov2005}. In this attack, Eve attempts to deterministically control Bob's basis choice and detection outcomes without terminating the protocol. To achieve this, Eve needs to maintain the expected detection rate between Alice and Bob, and keep the QBER below the termination threshold during her attack. In our practical attack model, we assume that Eve knows the attack angles for each polarization state, as well as the detection efficiency ratios between the detectors. Eve intercepts signals sent by Alice using an active basis choice receiver and superconducting nanowire detectors with an overall detection efficiency of $85\%$. This interception could be done right in front of Alice's setup, to negate the turbulence effect on Eve's measurement. She then generates a signal with the same polarization state as her measurement result, and sends it to Bob at the ideal attack angle. These fake signals may suffer from atmospheric turbulence in transmission to Bob.

We assume that Eve is restricted to today's technology, and uses a weak coherent state for her resend signal. Thus, Eve can control the mean photon number $\mu$ of her pulses, as well as mimic scintillation caused by turbulence in the free-space channel to avoid arousing suspicion. Several free-space QKD systems employ pointing and tracking systems that use a bright beacon source and wave front sensor \cite{bourgoin2015,pugh2017, yin2017} which could be adapted by Bob to monitor and correct beam wander. However, this pointing system uses a separate beacon laser at a different wavelength. This beacon laser does not need to be tampered with by Eve, and the pointing is unaffected by her attack. In the worst case, Eve could perform an intercept-and-resend attack on the beacon beam such that Bob's receiver is pointed according to her designated direction. Thus, this pointing and correction system cannot prevent the attack in our model.

To verify the possibility of a successful attack, we use an optimization program to find the mean photon number that Eve should use for each attack angle to match Bob's expected total detection probability while minimizing the QBER. Our detailed attack model and the optimization process are explained in Ref.~\onlinecite{sajeed2015a}. 

We first characterize a spatial mode attack for a channel without turbulence ($r_0 = \infty$) before considering a turbulent channel. The optical alignment between the sender (Alice) and the receiver (Bob) is optimized by equalizing the detection count rates of the four polarization channels for a beam propagating through the center of the scanning lens $\text{L}_\text{E}$ [i.e., along the green center ray shown in \cref{fig:setup}(a)]. This initial alignment represents normal operation which has a scanning angle $\phi = \theta = 0$. We then move the two-axis translation stage to adjust the position of lens $\text{L}_\text{E}$, and record the four detection efficiencies (\textbf{H}, \textbf{V}, \textbf{D}, and \textbf{A}) for different angles ($\theta, \phi$). In principle, the tip-tilt angles induced on the beam by the scanning lens are equivalent to including additional Zernike polynomial terms in the SLM hologram. Furthermore, the order in which the different Zernike polynomials are applied to the beam is interchangeable. As a result, our configuration of having the scanning lens follow the SLM is equivalent to Eve first steering the beam before it propagates through atmospheric turbulence. The scan is performed in $135\:\micro\radian$ steps, covering a range of $\pm 2.7~\milli\radian$, which corresponds to a lateral displacement of $\pm 35~\milli\meter$ along the front lens $\text{L1}$ of the QKD receiver. 

In order for an angle to be a valid attack angle for channel $k$ ($k = \bf{H}, \bf{V}, \bf{D},\:\mathrm{or}\:\bf{A}$), it must satisfy the condition that the probability of detection in channel $k$ is $\delta_k$ times greater than the detection probabilities of the two channels in the other basis. For example, if $k=\bf{H}$, then $\mathrm{min}\{\tau_H/\tau_D,\tau_H/\tau_A\}>\delta_H$, where $\tau_k$ is the normalized detection probability defined as the ratio between the detection rate at the attack angle over the expected detection probability of Bob. We continuously increase the threshold $\delta_k$ until only a few attack angles satisfy these conditions. From the attacker's point of view, it is desirable to have $\delta_k$ as large as possible because a large value means an increased chance that detector $k$ will click while minimizing the detection probabilities of the two other channels, which improves Eve's knowledge of Alice's state.

The scan results without turbulence ($r_0 = \infty$) for the four polarization channels are shown in \cref{fig:scan}(a), and the corresponding detection efficiency mismatch parameters are listed in \cref{tab:mismatch}. There are noticeable features that cause efficiency mismatch, such as the side peak visible below the center peak in the \textbf{H} detector's map, and the outer ring in all four detector maps. The valid attack angles for the \textbf{H} detector correspond to when the click probability is 22 times higher than {\bf D} and {\bf A} detectors (i.e.,\ $\delta_H = 22$), and the normalized detection probability $\tau_H = 0.1$. Although the mismatch ratios on {\bf D} ($\delta_D = 5$) and {\bf A} ($\delta_A = 1.2$) channels are small, the mismatch in {\bf H} and {\bf V} ($\delta_V = 30$) channels are sufficient for a successful attack under our assumption that Alice and Bob only monitor the total count rate (not individual channels). 

The optimized QBER as a function of transmission loss between Alice and Bob for a channel without turbulence is shown in \cref{fig:loss-QBER}. In a practical scenario, Alice and Bob might experience transmission efficiency fluctuations in their quantum channel. As a result, they need to tolerate some deviation in their key rate from their estimated value. Shown in \cref{fig:loss-QBER} is the QBER during Eve's attack as a function of the lowest transmission loss acceptable to Alice and Bob. In the next section, we examine the success of Eve's attack in the presence of turbulence.

\begin{table}
\caption{Detection efficiency mismatch parameters for attack data shown in \cref{fig:scan,fig:loss-QBER}. $\tau_k$ is the relative detection efficiency at an attack angle compared to the normal incidence case, and varies for different turbulence strengths due to changes in the scanning features that lead to valid attack angles. The value of the threshold of detection efficiency ratio $\delta_k$ decreases under stronger turbulence. If the $\delta_k$ are too low, it is impossible for Eve to find an optimal mean photon number for her resend signal that matches Bob's expected detection rate and does not induce error above the termination threshold. *~denotes the turbulence strengths where an attack is not feasible. \vspace{\baselineskip}}
\label{tab:mismatch}
\def\arraystretch{1.2}
\begin{tabular}{l|C{.7cm} C{.7cm} C{.7cm} C{.7cm}|C{.7cm} C{.7cm} C{.7cm} C{.8cm}}
\hline\hline
\multirow{2}{*}{\begin{tabular}[c] {@{}l@{}} $r_0\:(\centi\meter)$ \end{tabular}} & \multicolumn{4}{c|}{$\delta_k$} & \multicolumn{4}{c}{$\tau_k$} \\ \cline{2-9}  
            & H     & V    & D    & A   & H    & V     & D   & A   \\ \hline
$\infty$    & 22     & 30   & 5.0    & 1.2   & 0.1  & 0.03  & 0.3 & 0.001 \\ 
7.00        & 20     & 5.0   & 1.03    & 3.5  & 0.3 & 0.4   & 0.8 & 0.7 \\ 
3.50        & 8.0      & 2.5   & 1.08    & 2.3  & 0.5  & 0.15   & 0.85 & 0.5 \\ 
2.21      & 4.5      & 1.8    & 1.15  & 2.21  & 0.4  & 0.2   & 0.85 & 0.2 \\
1.53      & 3.0      & 2.0  & 1.7  & 1.25  & 0.45  & 0.3   & 0.85 & 0.02 \\  \hline
\rowcolor{lightgray}[1.97pt] 1.00*      & 1.2      & 1.7  & 1.02  & 1.01  & 0.25  & 0.4   & 0.3 & 0.15 \\ 
\hline\hline
\end{tabular}
\end{table} 

\section{Practical attack under turbulence}
\label{sec:attack}

To simulate our attack in the presence of atmospheric turbulence, we use a set of 29 holograms per turbulence strength, as described in Sec. II. We have performed scans of our QKD receiver for five different turbulence strengths: $r_0 = 7.00$, $3.50$, $2.21$, $1.53$, and $1.00~\centi\meter$. Our preliminary experiments that included tip-tilt wander caused by turbulence (i.e.,\ the second and third terms of Zernike polynomials) showed that if Eve does not correct for beam wander caused by turbulence, her attack is not feasible even under very weak turbulence ($r_0 = 7.00~\centi\meter$) corresponding to typical high-altitude atmospheric conditions. The beam wander from tip-tilt alone was a strong enough disturbance to significantly hinder her attack. We then repeated the attack under the assumption that Eve can correct for tip-tilt beam wander using adaptive optics, such as with a deformable mirror or SLM. These corrections are implemented in our scans by setting the weight of the second and third terms of Zernike polynomials to zero.

\begin{figure}
\includegraphics{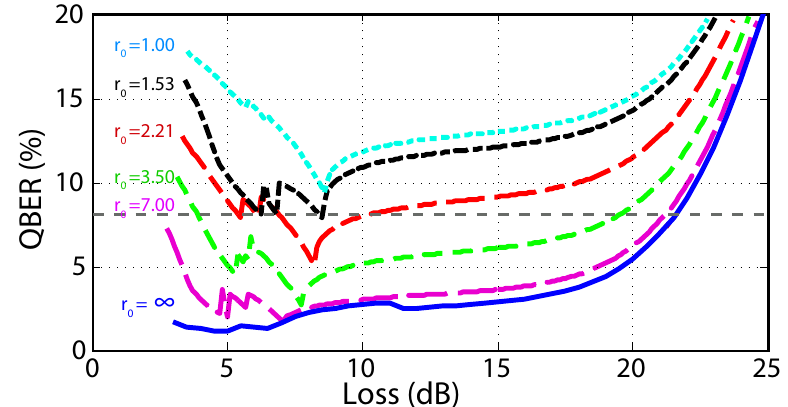}
\caption{Modeled attack performance. Quantum bit error rate (QBER) as a function of transmission loss for no turbulence (blue solid line) and different turbulence strengths corresponding to $r_0 = 7.00~\centi\meter$ (pink dashed line), $3.50~\centi\meter$ (green dotted line), $2.21~\centi\meter$ (red dot-dashed line), $1.53~\centi\meter$ (black dashed line), and  $1.00~\centi\meter$ (cyan dashed line). The horizontal gray dashed line denotes the $8\%$ threshold where Eve's attack is successful when QBER is below this value in our attack model. The maximum transmission loss where Eve's attack is successful decreases as turbulence strength increases. The mismatch ratios are too small in the case of $1.00~\centi\meter$ ($\delta_k \leq 2$ for all channels), and the optimization program could not find a solution with a QBER below 8\% threshold given any transmission loss. The higher QBER at low loss (i.e.,\ $3.5$--$7~\deci\bel$) is because Eve has to send higher mean photon number states for channels with lower $\delta_k$ in order to match the expected detection rate of Bob.}
\label{fig:loss-QBER}
\end{figure}

In order to maintain accuracy and stability in our scans, we have chosen to cycle through all 29 holograms at one lens position before moving the translation stage to the next position. This method ensures each hologram is applied to the same scanning angle. We then repeat this scanning process for a total of 1681 angle positions, and record 29 separate detection rates per attack angle for each of the four polarization channels. To represent the Gaussian distribution of centroid displacements discussed in \cref{sec:TurbEmu}, the final normalized detection efficiency of each detector $\tau_k$ is given by a weighted average of the detection rates from each hologram per scanning angle ($\theta,\phi$),
\begin{equation}
\tau_k(\theta,\phi) = \sum_{i=1}^{N} \Phi_i \tau_{k,i}(\theta,\phi),
\label{eq:efficiency-normalised}
\end{equation}
where $\tau_{k,i}$ is the average detection efficiency of the $k$ detector under the holograms selected from the $i$th radius. $\Phi_i$ is the probability of occurrence of the $i$th partition discussed in \cref{sec:TurbEmu}. $N = 5$ is the number of partitions. We select one sample hologram for no turbulence, eight samples each for $0.5\sigma$, $1\sigma$, $2\sigma$ partition, and four samples from the $3\sigma$ partition. The samples are given the weight factor corresponding to the radius from the sample used to the next larger sample, thus representing the best case hologram from this range. This weight factor ensures that the samples form an optimistic (easier to hack) representation of the turbulence effect, and therefore ensure that any turbulence found to not be vulnerable to attacks is indeed safe under the parameter monitoring assumptions. The total detection rate $\tau_k$ is used to find valid attack angles under turbulent conditions using the same method as without turbulence. We then repeat this process for different turbulence strengths from very weak ($r_0 = 7.00~\centi\meter$) to stronger turbulence emulating low-altitude sea level conditions ($r_0 = 1.00~\centi\meter$). A map of successful attack angles and the corresponding detection efficiency mismatch parameters are shown in \cref{fig:scan}(b)--(f) and \cref{tab:mismatch}.

Our scanning results in \cref{tab:mismatch} show that as the turbulence strength increases, the mismatch ratios $\delta_k$ are significantly reduced.  We can see in \cref{fig:scan} that the features that are responsible for efficiency mismatch become blurry and eventually disappear as turbulence increases in strength, and it becomes harder for Eve to maintain a precise attack angle when $r_0 \leq 1.53~\centi\meter$. For stronger turbulence ($r_0 = 1.00~\centi\meter$), the only remaining hackable feature is the displacement of the center peaks due to a slight misalignment between the fiber couplers in each arm of the receiver. As a result, most of the attack angles at stronger turbulence are found closer to the center peak. However, they do not result in a successful attack for $r_0 < 1.53~\centi\meter$ because the induced QBER is above the $8\%$ termination threshold. 

In order to perform a quantitative verification of an attack, we use an optimization program to find the minimal QBER as a function of transmission loss. The results in \cref{fig:loss-QBER} show that the optimized QBER for an attack in stronger turbulence ($r_0 = 2.21~\centi\meter$) is higher than that of weaker turbulence ($r_0 = 7.00~\centi\meter$). If we assume that the QBER threshold for Alice and Bob to terminate the protocol is $8\:\%$, then the attack without turbulence is successful as long as the transmission loss between Alice and Bob is less than $21\:\mathrm{dB}$, whereas in the presence of turbulence, Eve can successfully attack this receiver for $r_0 \geq 2.21~\centi\meter$ when the transmission loss is less than $10~\:\mathrm{dB}$ but higher than $7~\:\mathrm{dB}$. Using \cref{eq:sigma}, $r_0= 2.21~\centi\meter$ is equivalent to Eve having her resend setup approximately $0.5~\kilo\meter$ away from Bob's receiver in typical sea level turbulence conditions ($C_n^2 = 1.8\times10^{-14}~\meter^{-2/3}$). Eve is unable to match Bob's count rate for transmission loss below $3.5~\:\mathrm{dB}$ even if she uses all four channels due to Eve's non-perfect detection efficiency. Therefore, the optimization program could not find a solution matching Bob's total detection rates for transmission losses below $3.5~\deci\bel$.

The result for $r_0 = 1.53~\centi\meter$ shows that there is only a small loss window (around $8.5~\:\mathrm{dB}$) where Eve can attack without inducing a QBER higher than the threshold. Using \cref{eq:sigma} and the value of $C_n^2$ given above, this $r_0$ corresponds to a distance of $1~\kilo\meter$. At lower transmission loss (i.e.,\ $3.5$--$7~\deci\bel$), the expected detection rate at Bob is too high for Eve to match using a single channel, and therefore she must also use the other channels that have a lower $\delta_k$. This causes the QBER to increase and results in the irregularities seen for loss below $7~\deci\bel$ when the number of channels being used is changed. The QBER curves become smoother at higher loss once Eve can fully replicate Bob's detection rates while only sending signals to a single polarization channel, which takes advantage of the greatest efficiency mismatch for an optimized attack. The mismatch ratios in the case of $1.00~\centi\meter$ ($\delta_k \leq 2$ for all channels) are too small for the optimization program to find a solution for a QBER below the threshold given any transmission loss.

Implementations of QKD can and should monitor counts at each detector to ensure they remain relatively balanced. The higher QBER obtained when Eve is forced to send states to channels with lower mismatch ratios illustrates how monitoring each channel would increase the difficulty of a successful attack. However, it is uncommon in practice to monitor individual count rates, and there are no current standards or established guidelines for allowable variation in detection rates. The added constraint to maintain precise detection rates would make hacking more difficult for an eavesdropper, but does not in itself prevent an attack. It also does not invalidate the current work of determining if bounds exist on the turbulence strength where QKD systems can be hacked. 

\section{Theoretical limit of attack under turbulence}
\label{sec:theory}

The attack described in \cref{sec:attack} is only one particular example of an intercept-resend attack. Other attacks in this class may exist which shows that a QKD system with detection efficiency mismatch could be insecure if the security analysis does not take the mismatch into account. Whenever the observed and monitored data are compatible with an IR attack, no secret key can be obtained \cite{curty2004a,curty2005}.

For this reason, it is useful to ask the question whether the data we observe are consistent with an IR  attack or not. Along the way we can also answer the question whether a fine-grained analysis of the observations could exclude IR attacks, and thus potentially give a secure key where the coarse-grained analysis (which uses only average error rate and average detection rate) fails.

The handle to determine whether given data are compatible with an IR attack or not is the fact that IR attacks make the channel between Alice and Bob entanglement breaking. That is, this channel acting as one system of a bipartite entangled state will transform it into a separable bipartite state. So by verifying that the channel is not entanglement breaking, we can exclude the IR attacks. To do so, we do not require actual entanglement: we can probe the channel with non-orthogonal signal states, just as in any prepare-and-measure QKD set-up, and use the formalism of the source-replacement scheme (see for example Ref. \cite{ferenczi2012.PhysRevA-86-042327}) to formulate an equivalent thought set-up that virtually uses an entangled state.  The probabilities $p(a b| x y)$ between Alice's signal choice $a$ and Bob's measurement result $b$ for respective basis choices $x$ and $y$ can then be thought of as coming from measurements on this entangled state with both Alice and Bob performing measurements with POVM elements $M^{x,a}_A$ and $M^{y,b}_B$, respectively. If these observations serve as an entanglement witness, we have shown that the channel is not entanglement breaking. 

We can formulate the entanglement verification problem as the optimization problem
\begin{equation}
\label{eq:sdp}
\begin{array}{ll}
\text{find}& \rho_{AB} \\
\text{subject to} & \rho_{AB}\geq 0 \text{  and  } \rho_{AB}^{\Gamma_A}\geq 0 \\
                  & \text{Tr}(\rho_{AB}M_{A}^{x,a}\otimes M_B^{y,b})=p(ab|xy), \forall a,b,x y.
\end{array}
\end{equation}
Here $\Gamma_A$ is the partial transpose operation on Alice's system. If the above optimization problem is not feasible, then the state $\rho_{AB}$ is entangled \cite{peres1996}. In our previous work \cite{zhang2017}, we developed a method to solve the above optimization problem when detectors' 
efficiencies are mismatched and the dimension of the optical signal is unbounded.  

In this paper, we did not measure the joint distribution $p(ab|xy)$ of Alice and Bob directly in the experiment. However, given the characterization of detection efficiency mismatch from our experiment, we can deduce the joint distribution of Alice and Bob from the case without efficiency mismatch according to our simulation model. Using the method developed in Ref.~\onlinecite{zhang2017}, we found that when there is no turbulence or very weak turbulence $r_0 = 7.00~\centi\meter$, we cannot verify entanglement. Thus, the channel is vulnerable. This result is in agreement with the results in Ref.~\onlinecite{zhang2017}. 

However, when turbulence is stronger ($r_0 \leq 3.50~\centi\meter$), our calculation shows that entanglement can be verified. This means that there is no intercept-resend strategy for Eve that can match all of Alice and Bob's expected observations. This result is based on a strong condition where Eve needs to match all expected measurable parameters of Alice and Bob. Whereas, the results presented in \cref{sec:attack} were under the practical assumptions that Alice and Bob monitor only coarse-grained information, namely the total detection rate and error rate.  

\section{Conclusion}
\label{sec:conclusion}

We experimentally study how atmospheric turbulence in a free-space channel can affect an eavesdropper's ability to perform a spatial mode attack on a QKD receiver. We use a phase-only spatial light modulator to emulate atmospheric turbulence in the laboratory, the accuracy of which is verified by comparing measured far-field intensity distributions and centroid displacements to theoretical predictions. We then study a spatial mode detection efficiency mismatch attack under a range of atmospheric turbulence strengths to determine the maximum unsafe radius around the free-space QKD receiver. Our attack model is based on an intercept-resend attack under the practical assumptions that only the total detection rate and QBER are monitored by Alice and Bob. We find that for this particular receiver, an eavesdropper could attack a non-decoy state BB84 system from up to about $1~\kilo\meter$ away in typical sea level turbulence conditions ($r_0 = 1.53~\centi\meter$ for a $20~\centi\meter$ beam at $532~\nano\meter$). This result is assuming Eve can correct for basic tip-tilt beam wander using conventional adaptive optics. Eve's chances of success will be further reduced if Alice and Bob choose to monitor individual detection channel statistics. In this case, we theoretically find that an IR attack is still possible for weaker turbulence ($r_0 \geq 7.0~\centi\meter$). The assumption that an eavesdropper has physical limitations is not usually included in the security analysis of a QKD system. If there is a chance that Eve is inside this secure zone around Bob's receiver, or has advanced adaptive optics capacities to correct for beam aberrations, then extra care regarding these types of attacks may be required.

\begin{acknowledgments}
We thank Ben Davies and Brendon Higgins for assisting with our simulation code and data analysis. This work was funded by the US Office of Naval Research, Industry Canada, CFI, NSERC (Discovery program and CREATE project CryptoWorks21), Canadian Space Agency, Ontario MRIS, and the Ministry of Education and Science of Russia (program NTI center for quantum communications). P.C.\ was supported by Thai DPST scholarship. K.B.K.\ was supported by Canada First Research Excellence Fund. A.H.\ was supported by China Scholarship Council.

P.C.\ and K.B.K.\ contributed equally to this work and paper. 
\end{acknowledgments}

\def\bibsection{\medskip\begin{center}\rule{0.5\columnwidth}{.8pt}\end{center}\medskip} 
\bibliography{library}

\end{document}